\documentclass[fleqn,usenatbib]{mnras}

\usepackage{newtxtext,newtxmath}

\usepackage[T1]{fontenc}
\usepackage{svg}

\DeclareRobustCommand{\VAN}[3]{#2}
\let\VANthebibliography\thebibliography
\def\thebibliography{\DeclareRobustCommand{\VAN}[3]{##3}\VANthebibliography}

\usepackage{graphicx}	
\usepackage{amsmath}	

\title[Occurrence Studies for Various Planetary Types]{Constraining Planetary Formation Models Using Conditional Occurrences of Various Planet Types}

\author[Gajendran, Jiang, Yeh, and Sariya]{
Sridhar Gajendran$^{1}$,
Ing-Guey Jiang$^{1}$\thanks{jiang@phys.nthu.edu.tw},
Li-Chin Yeh$^{2}$, and Devesh P. Sariya$^{1}$
\\
$^{1}$Department of Physics and Institute of Astronomy, National Tsing Hua University, Hsin-Chu 30013, Taiwan\\
$^{2}$Institute of Computational and Modeling Science, National Tsing Hua University, Hsin-Chu 30013, Taiwan
}

\date{Accepted XXX. Received YYY; in original form ZZZ}

\pubyear{2015}

\begin{document}
\label{firstpage}
\pagerange{\pageref{firstpage}--\pageref{lastpage}}
\maketitle

\begin{abstract}
We report the conditional occurrences between three planetary types: super-Earths (\textit{m} sin \textit{i} $<$ 10 M{$_\oplus$}, P $<$ 100 days), warm Jupiters (\textit{m} sin \textit{i} $>$ 95 $M_{\oplus}$, 10 $<$ P $<$ 100 days), and cold Jupiters (\textit{m} sin \textit{i} $>$ 95 M{$_\oplus$}, P $>$ 400 days) for sun-like stars. We find that while the occurrence of cold Jupiters in systems with super-Earths is $22.2\substack{+8.3 \\ -5.4}$$\%$, compared to $10$$\%$ for the absolute occurrence rate of cold Jupiters, the occurrence of super-Earths in systems with cold Jupiters is $66.0\substack{+18.0 \\ -16.0}$$\%$, compared to $30$$\%$ for the absolute occurrence rate of super-Earths for sun-like stars. We find that the enhancement of super-Earths in systems with cold Jupiters is evident for sun-like stars, in agreement with several previous studies. We also conduct occurrence studies between warm Jupiters and super-Earths, and between warm Jupiters and cold Jupiters, to consolidate our methods. We conduct an independent observational test to study the effects of cold Jupiters against the inner multiplicity using the well-established giant planet host star metallicity correlation for all transiting planets found to date. The conditional occurrences we find here can be used to constrain the validity of various planetary formation models. The extremely interesting correlations between the super-Earths, cold Jupiters, and warm Jupiters can also be used to understand the formation histories of these planetary types. 
\end{abstract}




\section{Introduction}

The abundance of extrasolar planets discovered in recent years has created a renewed interest in planetary sciences. In particular, the multitude of extrasolar multiple planetary systems has given us an excellent opportunity to verify the validity of various planetary formation models. Even though planet formation is a complicated process involving various parameters, the continuity in the data of discovered exoplanets clearly implies that, in general, several common mechanisms play a part in the formation of generic planetary systems. The abundance of extrasolar multiple planetary systems has also given us an opportunity to explore the various planetary types and their effects in their respective planetary systems. One of the major discerning factors between different planetary formation models is their ability to explain the fraction of planets of different masses (i.e. different types) in a planetary system.

Super-Earths, a planet type whose formation is not well explained by any planetary formation models \citep{IdaLin2004, Mordasini2009, Morbidelli2016}, are nearly ubiquitous among the discovered planets. In particular, close-in or hot super-Earths, defined as a planet of around 1$-$10 M{$_\oplus$} or 1$-$4 R{$_\oplus$} and orbital period P $<$ 100 days \citep{Mayoretal2011, Fressin2013, Schlichting2018}, present a challenge to most planetary formation models. The two major models explaining their formation, namely, the in situ and inward migration models give rise to several contradicting conclusions. The numerical simulations reported in \citet{Izidoroetal2015} suggest that super-Earths are explained well by the inward migration model. However, their study also concludes that such planets will play a major role in defining the planetary composition of their respective systems and, in particular, would present an anti-correlation with giant planets. Consequently, \citet{Izidoroetal2015}, proposed that in systems with hot (close-in) super-Earths, the presence of gas giants should be rare. \cite{Schlaufman2014} used the metallicity scaling of giant planet occurrence with small planet occurrence rates to note that close-in multi-planetary systems suggest formation from high mass disks and that migration is important.

However, gas giants, particularly cold giants, are also known to be relatively abundant around sun-like stars. Using the ELODIE survey, \citet{Naefetal2005} first estimated a value of 7.5 $\pm$ 1.5$\%$ for the giant planet occurrence rate. \citet{Cummingetal2008} arrived at a value of 12.6 $\pm$ 1.6$\%$ from the Keck survey. Several follow-up studies \citep{Zechmeister2013, Wittenmyer2016, Wittenmyer2020, Fulton2021} found a value of $\approx$ 10$\%$ to be consistent for the occurrence rate of giant planets around sun-like stars. More recently, \citet{Bonomo2023} derived an occurence rate of $9.3 \substack{+7.7 \\ -2.9}$$\%$. Given their overwhelming numbers, it would appear highly unlikely that these two planetary types, super-Earths and gas giants, would be anti-correlated. Moreover, if the super-Earths really are dominant and prevent gas giants from forming in their systems, this would show up as a paucity of massive planets with short orbital periods. This, however, is not found to be true, as the period-ratio and mass-ratio of adjacent planets in multi-planet systems have been verified to present a positive correlation, suggesting that multi-planet systems with a large period ratio (and therefore a large orbital-radius ratio) also show a large ratio between their masses, with the more massive planet on the outside \citep{MazehZucker2003, Jiangetal2015}.

On the other hand, the in situ formation model for the origin of hot super-Earths produces an entirely opposite conclusion. For example, \citet{ChiangLaughlin2013} constructed a minimum-mass extrasolar nebula (MMEN) for explaining the formation of hot super-Earths and found that their MMEN is, in general, more massive (roughly six times) than the minimum-mass solar nebula (MMSN) \citep{Weidenschilling1977, Hayashi1981}. They were able to successfully explain several observed properties of the hot super-Earths, including their origin and their gas-to-rock fractions, well with the in situ model using their MMEN approach, but it also suggested that the more massive outer regions of the disk are more likely to form cold giants and therefore, these two planet populations must be correlated.

Our understanding of cold Jupiters (i.e., planets with mass $\gtrsim$ 0.3 M\textsubscript{J} and period $\gtrsim$ 1 AU) 
is much better than that of the super-Earths. The strong correlation between cold Jupiters and the metallicity of their host stars \citep{Santos2001, FischerValenti2005} is very well established. Their formation, characterized by three main phases, namely: (i) formation of the core by the accretion of planetesimals; (ii) slow accretion of both gas and solid material that is independent of time; and (iii) runaway gas accretion after the core reaches a critical mass \citep{Pollack1996}, has wide consensus.

In order to understand the origin of the enigmatic super-Earths, one can rely on observational evidence. Their bulk compositions, for example, can be constrained to a reasonable degree by the measurements of their mass and radius \citep{WuLithwick2013, Marcy2014, Hadden2017}. One can also study their dependence on host star metallicities to understand their formation and evolution pathways \citep{Mulders2015, WangandFischer2015, Zhuetal2016}. Alternatively, one can study the correlations between super-Earths and other planetary types to obtain information about the conditions required for their formation and their influence on adjacent planets. Since we have a strong theoretical and observational understanding of cold Jupiters, we can begin by looking for the correlations between these two planet types.

A comprehensive study of the relationship between super-Earths and cold Jupiters for sun-like stars was carried out by \citet{ZhuWu2018}. Their work concluded that while super-Earth hosts are approximately thrice as likely to have a cold Jupiter in them, cold Jupiter hosts were almost certainly ($\approx$ 90 $\pm$ 20$\%$) accompanied by super-Earths. This result implied that super-Earths and cold Jupiters do not compete for solid material in their formation stages. Such a correlation will help in establishing the necessary constraints on planet formation models for explaining super-Earth formation.

The high conditional probability of cold Jupiter hosts accompanied by super-Earths was independently studied again by \citet{Bryan2019}. \citet{Bryan2019} reported that $102\substack{+34 \\ -51}$$\%$ of stars with Jupiter analogs (3–7 AU, 0.3–13 M\textsubscript{J}) also host an inner super-Earth. Their study once again established that the presence of outer gas giants does not suppress the formation of inner super-Earths and that these
two populations of planets instead appear to be correlated.

Both studies concluded that nearly all Jupiter analogs host inner small planets (super-Earths), with uncertainties in their estimates of conditional probabilities due to the indirect nature of the Bayesian inference method they used. To directly estimate this fraction using observational data is difficult since a large sample of cold Jupiters with radial velocity (RV) data sets that are also sensitive to the presence of small inner planets is necessary. This requires long-baseline RV observations with high-cadence radial velocities or a good follow-up transit photometric survey. 

Recently, \citet{Rosenthal2022} made use of the California Legacy Survey  \citep{Rosenthal2021} to directly estimate this fraction. They concluded that while  $41\substack{+15 \\ -13}$$\%$ of systems with a close-in, small planet also host an outer giant, compared to $17.6\substack{+2.4 \\ -1.9}$$\%$ for stars irrespective of small planet presence, $42\substack{+17 \\ -13}$$\%$ of cold giant hosts
also host an inner small planet, compared to $27.6\substack{+5.8 \\ -4.8}$$\%$ of stars irrespective of cold giant presence. Their study concluded that the probability of hosting a small planet  given the presence of an outer gas giant is 1$\sigma$ enhanced over the absolute probability of hosting a small planet. Since this estimation is from direct measurements (as opposed to the indirect Bayesian inference of \citealp{ZhuWu2018} and \citealp{Bryan2019}), the correlation between super-Earths and cold Jupiters is strongly established once again.

In this study, the same line of reasoning as in \citet{ZhuWu2018} would be used to understand the super-Earth-cold Jupiter relationship. We will describe our samples and our method in Sec.~\ref{sec:sec2}. The conditional probabilities between the super-Earths, warm Jupiters, and cold Jupiters will be studied in Sec.~\ref{sec:sec3}. We will find the frequency of occurrence of cold Jupiters in systems with super-Earths and  use that to derive the frequency of super-Earths in systems with cold Jupiters in Sec.~\ref{sec:sec31} We will also analyze the dependence of the conditional probabilities on the metallicity of the hosts in Sec.~\ref{sec:sec31}. We will find the frequency of occurrence of warm Jupiters in systems with super-Earths and vice versa and compare how both conditional probabilities are related in Sec.~\ref{sec:sec32}. We will study the conditional probabilities between warm Jupiters and cold Jupiters in Sec.~\ref{sec:33}. These conditional probabilities are crucial to finding the necessary constraints one can place in coming up with a planetary formation model to explain these planetary types. 

In Sec.~\ref{sec:sec4}, 
the well-known giant planet host star metallicity relationship would be used. We will employ this result to conduct an observational test on the effects of giant planets on planet multiplicities for all transiting multiple planetary systems found to date. The implications of our conditional probabilities for the existing planetary formation models and a possible explanation of the origin of the super-Earths, in particular, will be discussed in Sec.~\ref{sec:sec5}.

\begin{figure*}
	\includegraphics[width= \textwidth]{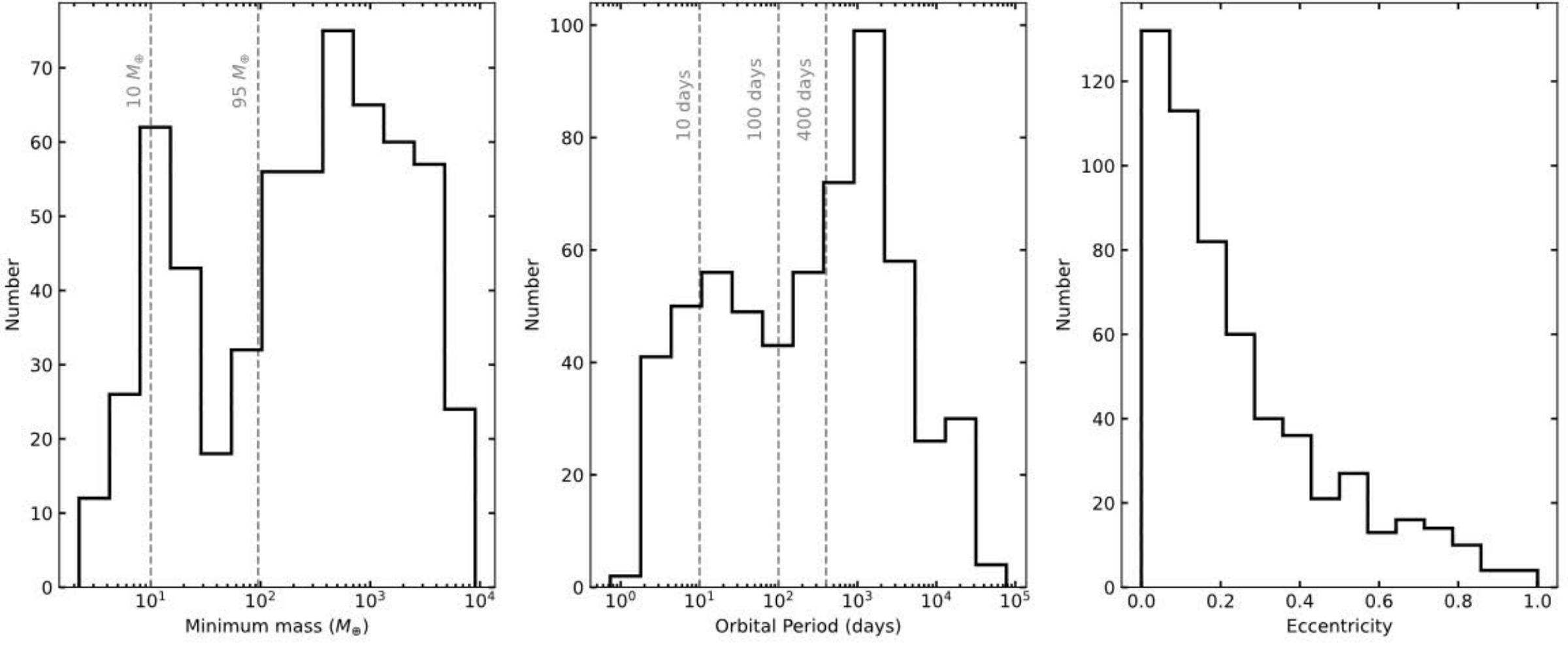}
    \caption{Histograms of masses, periods and eccentricities for all the RV planets in our samples that are known to have sun-like hosts. The dashed lines correspond to our definitions of super-Earths, warm Jupiters, and cold Jupiters. The two vertical lines in the leftmost panel, i.e., the mass histogram, correspond to our definitions for the masses of super-Earths and giant planets. The first two vertical lines in the middle panel, i.e., the period histogram, correspond to our definition of the periods of warm giants and super-Earths, and the third is placed between that of cold Jupiters.}
    \label{fig:fig1}
\end{figure*}

\begin{figure*}
	\includegraphics[width= 0.9\textwidth]{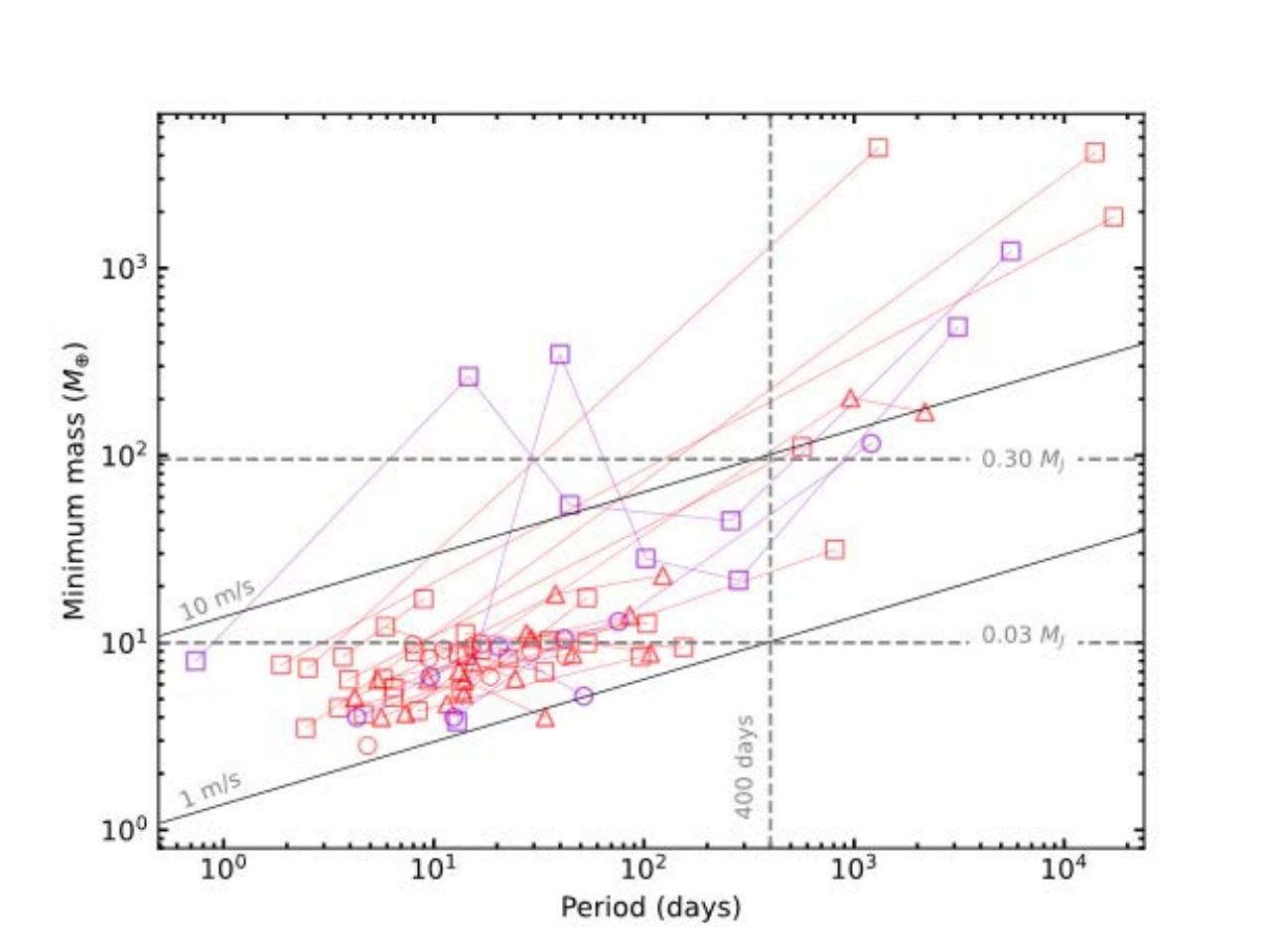}
    \caption{The period-mass plot of the planets in our sample. Systems with multiple planets (under the same host) are connected with solid lines. The two dashed horizontal lines correspond to our definitions for the masses of super-Earths and cold Jupiters and the vertical dashed line corresponds to our definition for the period of cold Jupiters. The two solid diagonal lines represent two characteristic RV semi-amplitudes of most current RV surveys. The figure implies that the RV data capable of detecting a super-Earth (in the lower left quadrant) in a multiple planetary systems must also reveal the presence of a cold Jupiter (in the upper right quadrant) in the same system (provided the RV time-spans are long enough).}
    \label{fig:fig2}
\end{figure*}

\section{The Data and Method}\label{sec:sec2}

The data set used for this study was obtained from the NASA Exoplanet Archive\footnote{http://exoplanetarchive.ipac.caltech.edu}
on December 1, 2023. Table~\ref{tab:table1} 
and Table ~\ref{tab:table2} show a section of our planets. For our first objective of calculating the conditional probabilities between the super-Earths and cold Jupiters, it is reasonable to use only the planetary systems discovered using the radial velocity (RV) method. This is due to the fact that for a given system, the RV data capable of revealing super-Earths is also most certainly capable of revealing cold Jupiters in that system, provided the RV time-spans are long enough (at least half the orbital period of the cold Jupiter) \citep{ZhuWu2018}. We further refine our data set by including only the detected planets with RV semi-amplitudes of \textit{K} $>$ 1 m s\textsuperscript{$-1$}. Any planet below this value presents a high refutation probability and is therefore discarded \citep{Rajpaul2016}.
For our next constraint, we choose planets with hosts that are sun-like, defined to be those with T\textsubscript{eff} in the range 4700$-$6500 K and log g $>$ 4.0. Therefore, all the constraints applied to our data set for calculating the conditional probabilities are summarized as follows:
\begin{enumerate}
	\item Planets must be detected by the radial velocity method 
 (with RV semi-amplitudes K $>$ 1 m s\textsuperscript{$-1$}).
	\item The hosts must be sun-like.
\end{enumerate}
The list of these extrasolar multi-planet systems is presented 
in Table~\ref{tab:table1}.
In addition, Figure~\ref{fig:fig1} shows the histograms of masses, periods, and eccentricities respectively, for all the RV planets from the NASA exoplanet archive that fit our requirements. The dashed lines correspond to our definitions of super-Earths (\textit{m} sin \textit{i} $<$ 10 M{$_\oplus$}, P $<$ 100 days), warm Jupiters (\textit{m} sin \textit{i} $>$ 95 $M_{\oplus}$, 10 $<$ P $<$ 100 days), and cold Jupiters (\textit{m} sin \textit{i} $>$ 95 M{$_\oplus$}, P $>$ 400 days). More particularly, the two vertical dashed lines in the leftmost panel of Figure~\ref{fig:fig1} (i.e., mass histogram) correspond to our definitions for the masses of super-Earths and giant planets, respectively. Similarly, the three vertical dashed lines in the middle panel (i.e., period histogram) correspond to our definitions for the periods of warm giants, super-Earths, and cold Jupiters. 

Figure~\ref{fig:fig2} explains our reasoning for choosing only the planets detected by the radial velocity method. All the multiple planetary systems in our sample that are known to host super-Earths are placed in the plot based on their masses and periods. The two dashed horizontal lines correspond to our definitions for the masses of super-Earths and cold Jupiters respectively and the vertical dashed line corresponds to the period threshold for the super-Earths and cold Jupiters. The two solid diagonal lines represent two characteristic RV semi-amplitudes of most current RV surveys. As the figure implies, for multiple planetary systems (and those with the same host star), the outer cold Jupiters induce RV semi-amplitudes greater than (or at least equal to) those by the super-Earths. Therefore, we can safely assume that the RV data capable of detecting a super-Earth in a multiple planetary system will also most certainly reveal the presence of a cold Jupiter in the same system, provided the system is observed for a long enough period of time. 

The conditional probabilities between the various planetary types are found using direct estimation from the data set and by using the Bayes theorem. The reasoning behind the choice of the respective method is explained in detail in the following section. The errors are calculated by first assuming a uniform prior and finding the posteriors from the likelihood that \textit{m} out of \textit{n} systems contain the desired planetary type as in \citet{Schlaufman2014} and \citet{Rosenthal2022}.

The second objective of this work is to study the effects of cold Jupiters on the inner multiplicity of planetary systems. We again use the NASA exoplanet archive 
and this time we consider only the planetary systems that are found by the transit method, irrespective of the host star type. 
Table~\ref{tab:table2} presents the list of these extrasolar multi-planet systems.

Although it is not yet possible to detect cold Jupiters for the majority of transit systems hosting small planets, we can resolve this issue by taking proxies. Since cold Jupiters are strongly correlated to the metallicity of the host stars, their impact on planet multiplicity must be more pronounced for systems around metal-rich stars. We further make the assumption that the transit multiplicity we find in our data set is the intrinsic multiplicity of these systems. To justify the reasoning behind this choice, we notice that systems found by the transit method are in general more likely to be multi-planetary  with a greater sensitivity to inner small planets than those found by the radial velocity method 
\citep{Escude2002, HolmanandNorman2005, Agoletal2005}. We then use the well-established giant planet host star metallicity correlation to infer the effects of cold Jupiters on the inner multiplicity of these systems. The effects of cold Jupiters on the inner multiplicity should show up as a paucity of high-multiple systems around metal-rich stars if we take the above correlation to be true.

\begin{table*}
	\centering
	\caption{List of extrasolar
multi-planet systems found using the radial velocity (RV) method that is used in this study for finding the different conditional probabilities. The RV timespans are obtained 
from The Data \& Analysis Center for Exoplanets (DACE) (\url{https://dace.unige.ch/}). Care is taken to ensure the RV timespans of all super-Earth hosts are known.}
	\label{tab:table1}
	\begin{tabular}{lccccr} 
		\hline
		ID & Host name & [Fe/H] & Period (Days) & Planet mass (M{$_\oplus$}) & RV time-span (Years)\\
		\hline
 1 & 14 Her b &  0.405 &   1765.0  & 2559.5 &  25.8	\\
  2 & 16 Cyg B b  &  0.06  &      798.5	& 565.7  & 32.8	\\
 3 & 30 Ari B b    & 0.19  &  335.1 &	4392.4 & \ldots	\\
 4&  47 UMa b &   0.04 &	 1078.0	  & 804.1  & 32.7	\\
 5 & 47 UMa c   & 0.04 &	 2391.0	 &  171.621 &  32.7	\\
6 & 47 UMa d &   0.04    &    14002.0	 &  521.2   & 32.7	\\
 7 &   51 Peg b &   0.18  &   4.2	& 146.2 &  25.3	\\
8 & 55 Cnc b &   0.35   &   14.7	& 264.0 &  25.3\\
9 &  55 Cnc c  &  0.35     & 44.4	&  54.5   & 25.3\\
 10 &  55 Cnc d    & 0.35  &     5574.2	& 1232.5  &  25.3\\
  11 & 55 Cnc e &    0.35  &  0.7	&     8.0 &  25.3\\
  12 & 55 Cnc f  &  0.35  &     259.9	 &  44.8   & 25.3\\
 13 & 61 Vir b  & $-$0.01 &	4.2	&      5.1  & 33.4\\
 14 & 61 Vir c &  $-$0.01   &    38.0	 &    18.2  & 33.4\\
    15 &    61 Vir d  & $-$0.01    &   123.0	&     22.9  & 33.4\\
   16 & BD+14 4559 b  &  0.17       &268.9& 330.5 &   3.5\\
17 &  BD+45 564 b  & $-$0.09    &   307.9	& 432.2  & \ldots\\
   18 &  BD+55 362 b   & 0.18     &  265.6  & 228.8 & \ldots\\
  19 & BD+63 1405 b  & $-$0.09    &  1198.4 & 3305.4 & \ldots\\
  20 & BD-00 4475 b  & $-$0.14	&723.2&	7961.6 & \ldots\\
 21 & BD-08 2823 b  & $-$0.06	&  5.6	 & 12.7  &  5.1\\
 22 &  BD-08 2823 c &  $-$0.07	& 237.6	  &    104.0  & 5.1\\
23 &  BD-10 3166 b   & 0.38  &    3.5	& 187.5   & 19.9\\
 24 &  BD-17 63 b  & $-$0.03	& 655.6	 &  1620.9   & 10.1\\
  25 &   CoRoT-20 c    & 0.14	& 1675.0	&  5403.1 & \ldots\\
   \ldots &  \ldots     &  \ldots  & \ldots & \ldots & \ldots \\	
   \ldots &  \ldots     &  \ldots  & \ldots & \ldots  & \ldots\\
 \hline
\multicolumn{4}{l}{\small Note: This table is available in its entirety in machine-readable form.}\\
	\end{tabular}
\end{table*}

\begin{table*}
	\centering
	\caption{List of extrasolar
multi-planet systems found using the transit method that is used in this study to study the dependence of planet multiplicity on host star metallicity.}
	\label{tab:table2}
	\begin{tabular}{lcccr} 
		\hline
		ID & Host name & [Fe/H] & Period (Days) & Planet mass (M{$_\oplus$})\\
		\hline
  	1 & AU Mic b  & 0.12 & 8.5  & 20.1 \\
         2& AU Mic c & 0.12  & 18.9 & 9.6 \\
       3 &  BD+20 594 b   & $-$0.15  & 41.7 & 22.2 \\
       4 & CoRoT-1 b & $-$0.3  & 1.5 & 327.4 \\
        5 &  CoRoT-10 b  & 0.26  & 13.2 & 874.0 \\
        6 &  CoRoT-11 b   & $-$0.03  & 2.9  & 740.5\\
        7 & CoRoT-12 b & 0.16  & 2.8 & 291.4  \\
       8 &  CoRoT-13 b  & 0.01 & 4.0  & 415.7 \\
	9 & CoRoT-14 b   & 0.05 & 1.5  & 2415.4\\
        10 & CoRoT-16 b   & 0.19  & 5.4  & 170.0 \\
        11 & CoRoT-17 b  & 0 & 3.8 & 772.2   \\
        12 &   CoRoT-18 b   & $-$0.1  & 1.9& 1102.8 \\
         13 &  CoRoT-19 b  & $-$0.02  & 3.9 & 352.8\\
       14 & CoRoT-2 b   & 0.04 & 1.9 & 1102.8   \\
       15 & CoRoT-20 b    & 0.14  & 9.2 & 1366.6\\
	16 & CoRoT-21 b  & 0  & 2.7 & 718.3 \\
	 17 & CoRoT-22 b & 0.17  & 9.7 & 12.2  \\		
	18 & CoRoT-23 b   & 0.05  & 3.6 & 890.0  \\	
	19 & CoRoT-24 b & 0.3  & 5.1 & 5.7\\
	20 & CoRoT-24 c     & 0.3  & 11.8 & 28.0 \\

     \ldots & \ldots     &  \ldots & \ldots & \ldots  \\	
    \ldots &  \ldots     &  \ldots & \ldots & \ldots \\
 \hline
\multicolumn{4}{l}{\small Note: This table is available in its entirety in machine-readable form.}\\
	\end{tabular}
\end{table*}

\begin{table*}
	\centering
 \caption{Various conditional probabilities of multi-planet systems based on the metallicity of the host stars.}
	\begin{tabular}{llllll} 
		\hline
		Conditional Probability & [Fe/H] < $-$0.2 & $-$0.2 $\leq$ [Fe/H] $\leq$ 0.2  & [Fe/H] > 0.2 & Combined & Method\\
		\hline
			\textit{P}(CJ|SE)  & $18.0\substack{+16.0 \\ -6.4}$$\%$ & $15.0\substack{+11.0 \\ -5.1}$$\%$ &  $50.0\substack{+17.0 \\ -17.0}$$\%$ & $22.2\substack{+8.3 \\ -5.4}$$\%$ & planet counting\\
   \textit{P}(SE|CJ)  &  $-$ &  $-$ & $-$   & $66.0\substack{+19.0 \\ -16.0}$$\%$ & Bayes theorem \\
    \hline
     \textit{P}(WJ|SE)  & $20.0\substack{+17.0 \\ -7.2}$$\%$ & $15.0\substack{+11.0 \\ -4.8}$$\%$ & $66.0\substack{+13.0 \\ -21.0}$$\%$ & $25.0\substack{+8.4 \\ -5.8}$$\%$  & planet counting\\
   \textit{P}(SE|WJ)  & $-$ & $-$ & $-$ &  $75.0\substack{+25.0 \\ -17.0}$$\%$  & Bayes theorem\\
   \hline
      \textit{P}(WJ|CJ)  & $13.0\substack{+10.0 \\ -4.2}$$\%$ & $13.6\substack{+3.6 \\ -2.5}$$\%$ & $18.5\substack{+5.1 \\ -3.5}$$\%$ & $15.9\substack{+2.7 \\ -2.1}$$\%$ & planet counting\\
   \textit{P}(CJ|WJ)  & $-$ & $-$ & $-$ & $15.9\substack{+2.7 \\ -2.1}$$\%$ & Bayes theorem \\

\hline
\end{tabular}

\label{tab:table3}
\end{table*}

\section{Conditional Probabilities}\label{sec:sec3}

\subsection{Conditional Probabilities of Super-Earths and Cold Jupiters}
\label{sec:sec31} 

We set our definition of a super-Earth to be a planet of mass, \textit{m} sin \textit{i} $<$ 10 M{$_\oplus$} and period, P $<$ 100 days in accordance with the most agreed definition for this class in exoplanet demographics \citep{Mayoretal2011, Fressin2013, Schlichting2018}. This introduces our next constraint and we choose only those systems that contain at least one such planet that fit our definition of a super-Earth. Single-planet systems are also included, removing them would further refine our sample. Figure~\ref{fig:fig2} shows that in our sample we find 36 multiple planetary systems that contain at least one super-Earth that fit our definition. 8 systems among them contain (one or more) cold Jupiters for sun-like stars. This frequency of cold Jupiters in the systems containing super-Earths is mathematically expressed as the conditional probability \textit{P}(CJ{$|$}SE), where CJ stands for cold Jupiters and SE stands for super-Earths respectively. One of the objectives of our study is to infer the inverse, i.e, the conditional probability of super-Earths in systems containing cold Jupiters, \textit{P}(SE{$|$}CJ). We will make use of the Bayesian inference to find this quantity and this will in turn reveal the role played by super-Earths in generic planetary systems. 

Given the nearly unity detection efficiency of cold Jupiters in systems with super-Earths (as explained in Sec.~\ref{sec:sec2}), this quantity is simply the fraction of systems with cold Jupiters. Table~\ref{tab:table1} also lists the RV time-spans of the multiple planetary systems detected by the RV method. We ensure that all super-Earth hosts exhibit RV time-spans exceeding four years, i.e., half the orbital period of the cold Jupiters. Therefore, a null detection almost certainly implies the absence of cold Jupiters within an 8-year orbit. Given these, we can be certain that the \textit{P}(CJ{$|$}SE), for our purposes, is simply given by the fraction of cold Jupiters in our sample of super-Earths. We estimate \textit{P}(CJ{$|$}SE) based on three divisions of the metallicity of the host stars. The results are tabulated in Table~\ref{tab:table3}. We find that \textit{P}(CJ{$|$}SE) is \textbf{$22.2\substack{+8.3 \\ -5.4}$$\%$} for solar-type stars for the combined sample, i.e., without any metallicity division. However, we find that this conditional probability has strong dependence on the metallicity of hosts and, in particular, it increases for high metallicities. This can be explained as the result of the well-known giant planet host star metallicity relationship. 

The absolute occurrence rate of cold Jupiters, \textit{P}(CJ), for sun-like stars could be found using the giant planet distribution function given in \citet{Cummingetal2008}. Using their giant planet distribution \citet{ZhuWu2018} arrived at a \textit{P}(CJ) of around 10$\%$ and our value of P(CJ{$|$}SE) 
 ($22.2\substack{+8.3 \\ -5.4}$$\%$) clearly indicates that cold Jupiters tend to appear more than twice as likely in systems containing super-Earths. This result is certainly intriguing as it shows that super-Earths and cold Jupiters can (and tend to) occur in the same system and that these two planet populations do not compete for solid material in their formation stages. This also reveals that these two planet types may not be anti-correlated after all, as explained by some planetary formation models.

In order to estimate the inverse, i.e., the occurrence rate of super-Earths in systems containing cold Jupiters, \textit{P}(SE{$|$}CJ), we could, in principle, use the same method as above. However, this may not be a reasonable approach to constraining this particular value for the following reasons:

\begin{enumerate}
	\item In systems containing cold Jupiters, the induction of RV semi-amplitude is much more dominated by the cold Jupiters in comparison to the super-Earths in the same system (refer Figure~\ref{fig:fig2}),
	\item Our current RV instruments are not capable (i.e., not sensitive enough) of detecting the majority of super-Earths.
\end{enumerate}

Long-baseline RV observations are required in order to directly determine the frequency of super-Earths in systems containing cold Jupiters. To be sensitive enough to determine the presence of super-Earths in such systems, we also require high-cadence and
high-precision RV observations \citep{Wittenmyer2009} or coverage by a photometric transit survey. These impose huge practical limitations in determining whether systems with cold Jupiters also contain super-Earths. Therefore, in order to find the conditional probability \textit{P}(SE{$|$}CJ), we can leverage our knowledge of the conditional probability \textit{P}(CJ{$|$}SE) and use the Bayes theorem to infer the former. The Bayes theorem, for our purposes, can be written as follows:

\begin{equation}
P(SE) \times P(CJ|SE) = P(CJ) \times P(SE|CJ)
\end{equation}

We have already established the values of \textit{P}(CJ) (10$\%$) and \textit{P}(CJ{$|$}SE) ($22.2\substack{+8.3 \\ -5.4}$$\%$) for sun-like stars. Finding the absolute occurrence rate of super-Earths, \textit{P}(SE), for sun-like stars is slightly complicated. This is because it is always easier to constrain the number of planets per star  than the frequency of stars that may contain planets as a whole. However, as determined by \citet{Zhuetal2018}, the fraction of sun-like stars with \textit{Kepler}-like planets (defined to be planets of R\textsubscript{P} $\geq$ R{$_\oplus$} and P $<$ 400 days), $\eta$\textsubscript{Kepler}, gives us a \textit{P}(SE) of $\approx$ 30$\%$. Even though this particular number only holds good for \textit{Kepler} planets and our study involves RV planets, we can still hold it as a good approximation due to the fact that the occurrence of super-Earths does not depend strongly on stellar metallicities \citep{Udryetal2006, Buchhaveetal2012, WangandFischer2015, Zhuetal2016}. With this effect of incompleteness,
the conditional probability derived here can be considered as a lower limit.

We arrive at a \textit{P}(SE{$|$}CJ) of $66.0\substack{+19.0 \\ -16.0}$$\%$ using our estimation from the Bayes theorem. This also tells us that systems with cold Jupiters are also most likely to have super-Earths. The two conditional probabilities we have calculated so far, \textit{P}(CJ{$|$}SE) and \textit{P}(SE{$|$}CJ), clearly indicate that the two planet populations are strongly correlated. This also clearly establishes the constraints that one may place on theories explaining the formation of super-Earths. As discussed previously, if super-Earths are well explained by a model supporting migration, then the observed correlation could not be true. We can therefore safely ascertain that even though migration scenarios work best in explaining hot Jupiters, e.g., \citep{WuLithwick2011, ValsecchiRosio2014, Attiaetal2021}, they may not be sufficient in explaining super-Earths.

The strong correlation that we found here also leads to another interesting consequence. We find that systems with cold Jupiters but no super-Earths, such as our own solar system, must be very rare. We can find this probability, \textit{P}(no-SE, CJ), using our knowledge of the conditional probability of super-Earths in systems containing cold Jupiters (\textit{P}(SE{$|$}CJ)) and the absolute occurrence rate of cold Jupiters (\textit{P}(CJ)), using the following relation:

\begin{equation}
P(no-SE,\ CJ) = [1 - P(SE|CJ)] \times P(CJ)
\end{equation}

With our values of \textit{P}(SE{$|$}CJ) (66.0$\%$) and \textit{P}(CJ) (10$\%$), we get a \textit{P}(no-SE, CJ) of 3$\%$. This establishes that while only 10$\%$ of sun-like stars contain cold Jupiters, an even lesser fraction of them contain cold Jupiters with no super-Earths. This implies that systems such as our own with cold Jupiters but no super-Earths must be rare. Interestingly, our result is consistent with \cite{Wittenmyer2011} which found that about 3.3$\%$ of stars hosted Jupiter-like cold giants with no inner giants.

We must also note that the value of  \textit{P}(SE{$|$}CJ) should have an upper limit because of some exceptional cases, like systems with hot Jupiters and solar system analogs; however, this cannot be constrained from our sample size, but we can make some estimates. It is known that planetary systems that host hot Jupiters do not usually have a small close-by companion like a super-Earth \citep{Steffen2012, Becker2015}. However, it is estimated that $\approx$ 50$\%$ of hot Jupiters have cold Jupiter companions \citep{Knutson2014}, the  absolute occurrence rate for hot Jupiters, \textit{P}(HJ), is estimated to be $\approx$ 1$\%$ \citep{Wright2012}  and this gives a \textit{P}(HJ{$|$}CJ) of $\approx$ 5$\%$. Considering this value and other reductions by systems like solar system analogs, we can estimate the upper limit of \textit{P}(SE{$|$}CJ) to be $\lessapprox$ 95$\%$. 

\subsection{Conditional Probabilities of Warm Jupiters and Super-Earths}
\label{sec:sec32}

We repeat the study to turn our attention towards warm Jupiters and super-Earths. The planets are selected in the same way as mentioned in Sec.~\ref{sec:sec2} and a warm Jupiter is defined to be a planet of mass, \textit{m} sin \textit{i} $>$ 95 $M_{\oplus}$ and period, 10 $<$ P $<$ 100 days. We find that the number of warm Jupiters in systems with super-Earths for sun-like hosts to be 9 out of 36 systems. This corresponds to a conditional probability \textit{P}(WJ{$|$}SE) of \textbf{$25.0\substack{+8.4 \\ -5.8}$$\%$}. We estimate this value once again under three divisions of metallicities of the hosts and the results are tabulated in Table~\ref{tab:table3}. As with \textit{P}(CJ{$|$}SE) we notice the dependence of \textit{P}(WJ{$|$}SE) to the metallicity of the hosts, i.e. increasing for high metallicity hosts. This dependence can be explained based on the well established giant planet host star metallicity correlation. The conditional probability of finding super-Earths in systems with warm Jupiters, \textit{P}(SE{$|$}WJ), cannot be constrained from the data directly for the same reason as explained in the previous subsection. We, therefore, resort to the Bayes theorem to estimate \textit{P}(SE{$|$}WJ). The Bayes theorem, for this purpose, is written as follows: 

\begin{equation}
P(SE) \times P(WJ|SE) = P(WJ) \times P(SE|WJ)
\end{equation}

Using the giant planet distribution function given in \citet{Cummingetal2008}, we can arrive at an estimate of 10$\%$ for the absolute occurrence rate of warm Jupiters, \textit{P}(WJ), for sun-like stars. Taking the absolute occurrence rate of super-Earths to be 30$\%$ (as in the previous subsection) and the conditional probability, \textit{P}(WJ{$|$}SE) to be $25.0\substack{+8.4 \\ -5.8}$$\%$, yields a value of $75.0 \substack{+25.0 \\ -17.0}$$\%$ for the conditional probability \textit{P}(SE{$|$}WJ). This high value once again shows that the prevalence of super-Earths in systems containing giant planets is very significant. 

\subsection{Conditional Probabilities of Warm Jupiters and Cold Jupiters}\label{sec:33}

Using the same definition for warm Jupiters and cold Jupiters, as before we constrain the conditional probabilities \textit{P}(WJ{$|$}CJ) and \textit{P}(CJ{$|$}WJ). We also select only the planets detected by the RV method and find that for solar-type stars (as defined in Sec.~\ref{sec:sec2}), out of 232 systems containing cold Jupiters, 32 also contain warm Jupiters, and the conditional probability of warm Jupiters in systems containing cold Jupiters, \textit{P}(WJ{$|$}CJ), therefore is  $15.9\substack{+2.7 \\ -2.1}$$\%$ for the combined sample, i.e., without any metallicity division. From Sec.~\ref{sec:sec31}, we know that the absolute occurrence rate of cold Jupiters for solar-type stars, \textit{P}(CJ), is  10$\%$ and our estimate of \textit{P}(WJ{$|$}CJ) implies that the occurrence of warm Jupiters is slightly enriched in systems containing cold Jupiters. We also notice from Table~\ref{tab:table3} that this conditional probability has a dependence on the metallicity of the hosts and increases for high metallicities. 

Even though we can directly estimate the conditional probability, \textit{P}(CJ{$|$}WJ), from the data similar to  \textit{P}(WJ{$|$}CJ) in principle, we use the Bayesian inference method (similar to the previous subsection). The Bayes theorem, can be re-written as follows:

\begin{equation}
P(CJ) \times P(WJ|CJ) = P(WJ) \times P(CJ|WJ)
\end{equation}

We find that with a \textit{P}(WJ) of 10$\%$ \citep{Cummingetal2008}, \textit{P}(CJ) of 10$\%$ and \textit{P}(WJ{$|$}CJ) of $15.9\substack{+2.7 \\ -2.1}$$\%$, we arrive at a value of \textit{P}(CJ{$|$}WJ) of $15.9\substack{+2.7 \\ -2.1}$$\%$ for sun-like stars. Since the absolute probabilities of \textit{P}(WJ) and \textit{P}(CJ) are equal for sun-like stars, we end up with the same value for both the conditional probabilities, \textit{P}(WJ{$|$}CJ) and \textit{P}(CJ{$|$}WJ).

We also find from the previous subsection that constraining \textit{P}(SE{$|$}CJ) directly from the data yields a value of $12.1\substack{+8.9 \\ -3.5}$$\%$ which clearly implies that the actual number of super-Earths in systems containing cold Jupiters cannot be obtained using the same method used for constraining the number of cold Jupiters in systems containing super-Earths. Therefore, we can be assured that our estimate of \textit{P}(SE{$|$}CJ) from the Bayesian inference must be a reasonable approach from our estimate of \textit{P}(WJ{$|$}CJ).

\begin{figure*}
	\includegraphics[width=\textwidth]{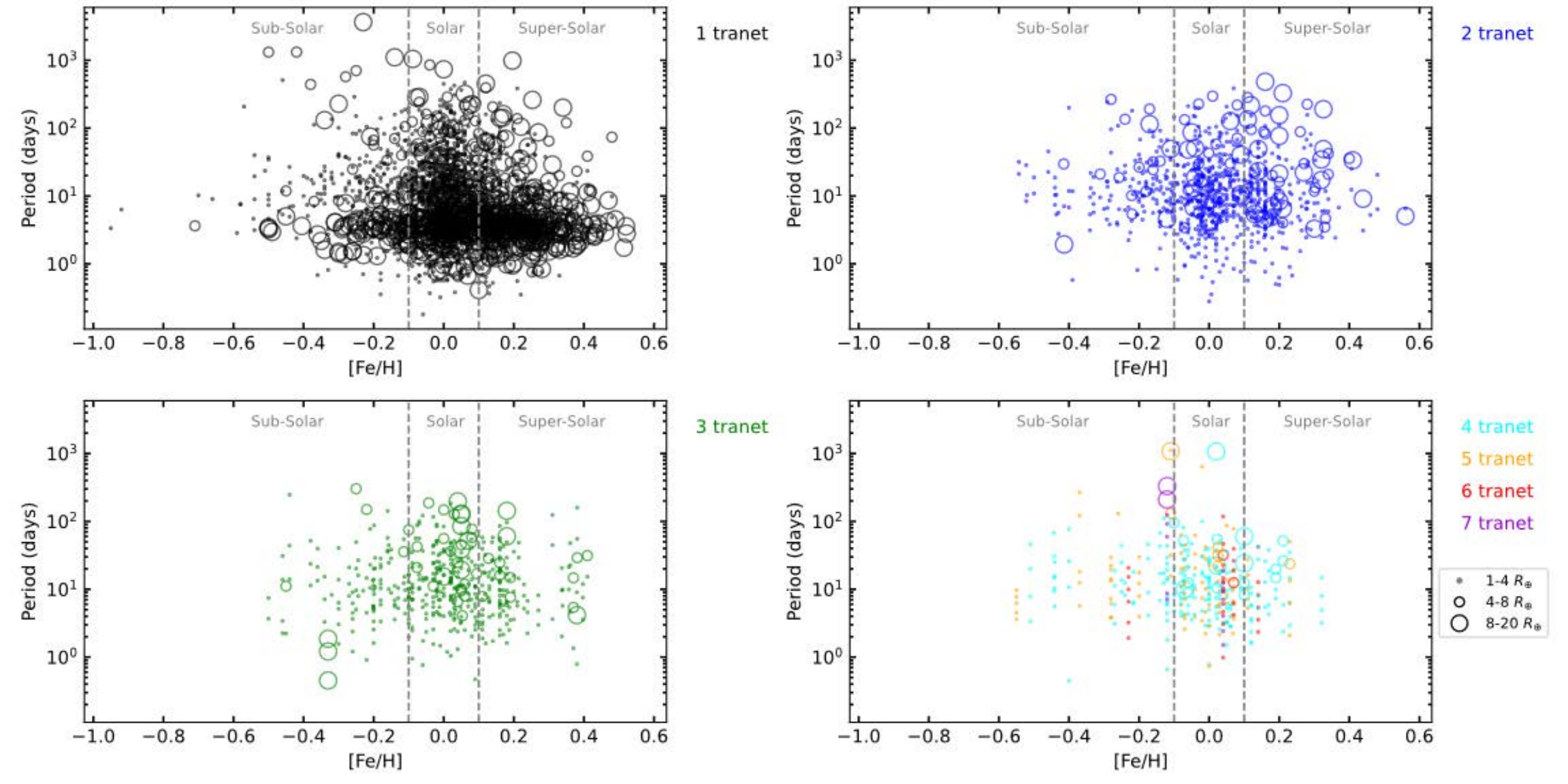}
    \caption{The plots of stellar metallicity ([Fe/H]) versus planetary orbital period for all transiting planets from the NASA exoplanet archive. Tranet stands for transiting planet. The transiting planets are split into four groups based on the multiplicity of their systems and are placed on four different panels. The two dotted vertical lines split the metallicity range into three metallicity regions, namely, sub-solar, solar, and super-solar. While low-multiplicity systems are evenly distributed across hosts of all metallicities there is a paucity of high-multiplicity systems around hosts of high metallicity indicating the effect of cold Jupiters on the inner planet multiplicity.}
    \label{fig:fig4}
\end{figure*}

\section{Planet Multiplicity versus Metallicity}\label{sec:sec4}

We now turn to the second objective of this study. Cold Jupiters are known to have significant eccentricities (e $\approx$ 0.3). This indicates that they have a strong tendency to disturb the inner planetary system if they are part of a multiple planetary system \citep{Matsumura2013, BeckerAdams2017, Hansen2017, Huang2017, Mustill2017, PuLai2018}. We can therefore expect a paucity of heavily-packed multi-planet systems around hosts that contain cold Jupiters. As a result, systems with cold Jupiters should have on average fewer planets within 400-day orbits than the ones without cold Jupiters.

On the other hand, while the conditional probability \textit{P}(CJ{$|$}SE) is $\approx$ $22.2\substack{+8.3 \\ -5.4}$$\%$ for our sample, i.e., for sun-like stars, it increases significantly for metal-rich systems  ($\approx$ $50.0\substack{+17.0 \\ -17.0}$$\%$). This is because of the well-established correlation between cold Jupiters and host star metallicity \citep{FischerValenti2005, Johnson2010, Petigura2018}. If cold Jupiters cause significant disturbance in the inner planetary systems, it will be seen as a paucity in the inner multiplicity in metal-rich systems. Therefore, one expects an anti-correlation between the multiplicity of the inner planetary systems and stellar metallicity.

We use data from the NASA exoplanet archive, and this time we only select those systems that are detected by the transit method, as opposed to the planets in the previous section, which only contained RV systems. This is both due to the fact that the systems found by the transit method are more in number, leading to a more meaningful statistic, and because they are more likely to be multi-planetary with a greater sensitivity to inner small planets than those found by the RV method \citep{Escude2002, HolmanandNorman2005, Agoletal2005}. 

We separated the transiting multiple-planetary systems (referred to as Tranet) into four groups based on their multiplicity ($k$), more specifically, into low-multiplicity ($k \leq 3$) systems with 1$-$Tranet, 2$-$Tranet and 3$-$Tranet each and high-multiplicity ($k \geq 4)$ systems with 4, 5, 6, 7$-$Tranet groups. Figure~\ref{fig:fig4} shows the planets in each group which are plotted based on their periods and the metallicities of their host stars. We further divide the planets in each group into three categories based on the metallicity of the host stars. Those planets with host stars  having metallicities [Fe/H] < $-$0.1 are labelled sub-solar; those with $-$0.1 $\leq$ [Fe/H] $\leq$ 0.1 are labelled solar, and those with [Fe/H] > 0.1 are labelled super-solar \citep{NissenEtal2020}. The dotted vertical lines mark our division of the metallicity range into sub-solar, solar and super-solar regions. The markers roughly indicate the sizes of the planets in each system. As Figure~\ref{fig:fig4} indicates, the outermost planets in most high-multiple systems have, roughly, orbital periods $P \geq 100$ days, making them vulnerable to outer giant planets.

From the figure, one also notices that systems with low-multiplicity tend to appear roughly equally around all metallicity values. Of the 2307 single tranets, 377 have sub-solar type hosts, 1329 have solar type hosts, and 601 super-solar hosts. For 2 tranets, out of 380 systems, 79 have sub-solar hosts, 185 have solar hosts, and 116 have super-solar hosts. For 3 tranets, the 157 systems are distributed such that 32 have sub-solar hosts, 59 have solar hosts, and 40 have super-solar hosts.  However, there is a clear paucity of high-multiplicity systems around metal-rich stars. For the 80 high-multiplicity systems we find in metal-poor stars, we expect to see roughly 38 of them in high-metallicity regime. But we only find 10 such systems and all these systems have all their planets with low orbital periods. Thus we may conclude that high-multiple systems around metal-rich stars can only survive provided that their planets all have low periods so that they can escape from the disturbances caused by the outer cold Jupiters \citep{LaiPu2017}. This deficit of high multiple systems around metal-rich stars can therefore be reasoned as an explanation for the suspected anti-correlation between multiplicity of the inner systems and stellar metallicities caused by the presence of cold Jupiters if we make the further assumption that the observed transit multiplicity for these systems are their intrinsic multiplicities.

\section{Conclusions}\label{sec:sec5}

The correlations between small, close-in planets and outer giant companions have huge implications for planet formation models. It has been proposed that outer cold giants may suppress the formation of inner small planets \citep{Lin1986, Moriarty2015, Ormel2017}. Warm or close-in giant planets may also have a similar effect \citep{Bitsch2020, Schlecker2021}, and these imply a population of small, close-in planets that are anti-correlated with both cold and warm gas giants. However, it is also speculated that the same conditions that aid the formation of giant planets, such as the high metallicity of the hosts, may also aid the formation of small planets. The conditional probabilities we estimated in this work from all the known exoplanetary systems give us a way to answer these contradictions and explain the formation histories of these various planet types.

Firstly, the conditional probabilities involving super-Earths and giant planets (warm or cold giants) show a dependence on the metallicity of the host stars as calculated by our three metallicity divisions. The marked increase in the conditional probability for these planetary types for high metallicity stars can be understood as resulting from the well-known host star-giant planet metallicity correlation. Therefore, our study serves as an indirect observational test for re-establishing this well-known correlation, i.e., in establishing that the occurrence rate of giant planets is heavily correlated to the metallicity of their host stars.

We now turn to the implications of our conditional probabilities for the formation scenario of various planet types in general and super-Earths in particular. The strong correlation between super-Earths and cold Jupiters continues to hold true for sun-like stars, in accordance with several previous studies \citep{ZhuWu2018, Bryan2019, Rosenthal2022}. This, taken at its face value, helps us to safely eliminate planetary formation models that propose an anti-correlation between these two planet types, in particular those involving large-scale migration scenarios, to explain the formation of super-Earths \citep{Izidoroetal2015, Izidoro2017}. Therefore, while the migration models may be suitable for explaining the formation of hot Jupiters, super-Earth formation, on the other hand, seems to favor in situ formation models. Since cold Jupiters are also explained well by the core accretion model, which favors in situ formation, the observed correlation between cold Jupiters and super-Earths seems natural.

Another significant result is about the prevalence of solar system analogs, i.e., the conditional probability of having cold Jupiters but no super-Earths. Our results imply that solar system analogs tend to be rare ($\approx$ 3$\%$). 
Previously, \cite{Schlaufman2014} 
proposed that only 0.06$\%$ of solar-type stars could host solar system analogs, i.e. systems hosting both a terrestrial and a long-period giant planet, in the Galaxy. 

We also find that while cold Jupiters are strongly correlated with super-Earths, they still have a strong influence on the multiplicity of planetary systems. Our results help establish that the systems with cold Jupiters must typically have a lower intrinsic multiplicity. Because of the well-established giant planet host star metallicity relationship, this also means that heavily packed systems appear less often around metal-rich stars than they do around metal-poor ones.

\section*{Acknowledgements}
We are grateful to the anonymous referee for good suggestions. 
This project is supported in part 
by the National Science and Technology Council, Taiwan, under
Ing-Guey Jiang's Grant MOST 111-2112-M-007-035 and
Li-Chin Yeh's Grant MOST 111-2115-M-007-008. 
This research has made use of the NASA Exoplanet Archive, which is operated by the California Institute of Technology, under contract with the National Aeronautics and Space Administration under the Exoplanet Exploration Program. This publication makes use of The Data \& Analysis Center for Exoplanets (DACE), which is a facility based at the University of Geneva (CH) dedicated to extrasolar planets data visualisation, exchange and analysis. DACE is a platform of the Swiss National Centre of Competence in Research (NCCR) PlanetS, federating the Swiss expertise in Exoplanet research. The DACE platform is available at https://dace.unige.ch.

\section*{Data Availability}

The dataset used for this study is obtained from the NASA exoplanet archive (\url{http://exoplanetarchive.ipac.caltech.edu}), and The Data \& Analysis Center for Exoplanets (DACE) (\url{https://dace.unige.ch}). The individual tables with the samples used for occurrence studies is available in its entirety in machine-readable form.



\bibliographystyle{mnras}
\bibliography{planet} 








\bsp	
\label{lastpage}
\end{document}